# Properties of the $Z(3)$ Interface in (2+1)-D $SU(3)$ Gauge Theory

S.T. West* and J.F. Wheater[a]

[a]Theoretical Physics, University of Oxford, 1, Keble Road, Oxford OX1 3NP, England, U.K.

A study is made of some properties of this interface in the SU(3) pure gauge theory in 2+1 dimensions. At high temperatures, the interface tension is measured and shows agreement with the perturbative prediction. Near the critical temperature, the behaviour of the interface is examined, and its fluctuations compared to a scalar field theory model.

## 1. Preliminaries

The pure glue sector of QCD, $SU(3)$ gauge theory, possesses several phases at finite temperatures ($T$). At low $T$, it exists in a single disordered phase where the vacuum is symmetric under the centre of the group, $Z(3)$, and the colour charge is confined; at high $T$, there are three distinct ordered, colour-non-confining, $Z(3)$-breaking phases, each degenerate vacuum corresponding to an element of $Z(3)$. At some critical temperature, $T_c$, there is a phase transition between the two regimes. Thus, two types of phase interface can occur in the theory: one between ordered and disordered phases, stable only at $T_c$, and one between two ordered phases, stable for $T > T_c$. The latter is often referred to as an "order-order" or "$Z(3)$" interface.

In the Euclidean formalism, the "time" dimension runs from 0 to $1/T$, and an order parameter for the phases is the Polyakov line, a time-ordered Wilson loop which wraps around the time boundary for a fixed spatial location $\vec{r}$:

$$L(\vec{r}) = \frac{1}{3}\mathrm{Tr}\mathcal{T}e^{ig\int_0^{1/T} d\tau A_0(\tau,\vec{r})}. \tag{1}$$

The expectation value of this Polyakov line gives the self-energy of a single quark in the gluon medium: $< L >= e^{-F/T}$. Confinement implies $F = \infty$ so that $< L >= 0$ in the disordered phase. For each ordered phase, $< L >$ is equal to a different element of $Z(3)$[1]. Unlike topologically trivial Wilson loops, then, the Polyakov line is aware of the phase structure of an $SU(3)$ system and can map out the structure of any interface.

It is widely believed, and has been shown by simulation[2], that the interfaces of $SU(3)$ show "critical wetting", meaning that a $Z(3)$ interface should be thought of as two order-disorder interfaces stuck back-to-back, leaving a thin slice of disorder within. Possible cosmological implications have been discussed in the past[3], with $Z(3)$ interfaces from the hot quark-gluon plasma splitting apart, as the early universe cooled to $T_c$, to spread confined matter outwards. However, questions have been raised as to whether these essentially Euclidean objects can have any significance in the real universe, especially as the high-$T$ order leads to unphysical thermodynamic properties for the interfaces[4] in the presence of fermions, and it has been argued that there may be only one true $Z(3)$ phase even at high $T$[5].

## 2. The Interface Tension at High $T$

The calculation of the interface tension, $\alpha$, is an instanton problem, in which the Polyakov line interpolates between different $Z(3)$ vacua. An effective one-dimensional theory describes the interface profile, with the effective action for the Polyakov line consisting of a classical kinetic part and a potential term formed by integrating out one-loop quantum fluctuations about a constant background field. The potential incorporates the $Z(3)$ symmetry, having minima when $L \in Z(3)$[1]. The instanton calculation has been performed[6], predicting a behaviour in 2+1 Euclidean dimensions of the form

$$\alpha = \alpha_0 T^{2.5}/g, \tag{2}$$

*Speaker at conference; funded by PPARC

where $\alpha_0$ is a constant predicted to be 8.33 for $SU(3)$ in the continuum, in the limit as $T \to \infty$. On the lattice, this prediction is modified to 9.82. It has also been argued[5] that higher-order infrared divergences invalidate this calculation, leaving only one true $Z(3)$ phase and, consequently, an interface tension of zero.

In order to test this prediction, we simulate $SU(3)$ pure gauge theory in 2+1 dimensions on a cubic Euclidean lattice of dimension $L_z \times L_x \times L_t$, $L_z > L_x \gg L_t$. The familiar Wilson plaquette action is used, with the variables defined on the lattice links being $SU(3)$ matrices. A "twist" [2] ensures the appearance of a physical interface: a set of plaquettes pierced by a skewer in the $x$ direction has its contribution to the action multiplied by an element of $Z(3)$, representing an effective change of variables; periodic boundary conditions then require the field to interpolate between the two resulting vacua elsewhere in the $z$ direction.

Simulating with and without the twist and measuring the average action in each case gives us the difference $\Delta S$, from which we obtain $\Delta F$, the free energy of the interface. With lattice spacing $a$, the interface "area" is $A = L_x a$ and the temperature $T = 1/(L_t a)$, so that

$$\frac{1}{\beta}\Delta S = \frac{\partial}{\partial\beta}\left(\frac{\alpha A}{T}\right) = \frac{L_x}{2\sqrt{6\beta}L_t^{1.5}}\alpha_0, \qquad (3)$$

where the second equality uses (2) and introduces the lattice parameter $\beta = 6/(ag^2)$.

Our simulations were performed for various $16.0 < \beta < 200.0$ with $L_t = 2$. The spatial dimensions were scaled with the interface width, $\sim \sqrt{\beta}$; for $\beta = 200$, we used $L_z = 96$ and $L_x = 32$. To estimate $\alpha_0$, it is necessary to extrapolate to $\beta = \infty$. Unfortunately, the exact form of the perturbative correction to $\alpha$ in (2) is not known. In 3+1 dimensions, a $1/\beta$ dependence is seen, but there is good reason to expect additional factors of $\ln\beta$ from the infrared divergences in 2+1 dimensions, as seen in the Debye mass[7]. In fig. 1, we perform the extrapolation using a best fit to a power of $1/\beta$, which we measure to be 0.85(10). The extrapolated value obtained is $\alpha_0 = 9.91(1)(14)$, the second error being associated with the power of $1/\beta$. This value is in good agreement with the instanton calculation, suggesting that the $Z(3)$ phases are indeed distinct at high $T$, and must be taken into account in calculations from the Euclidean path integral.

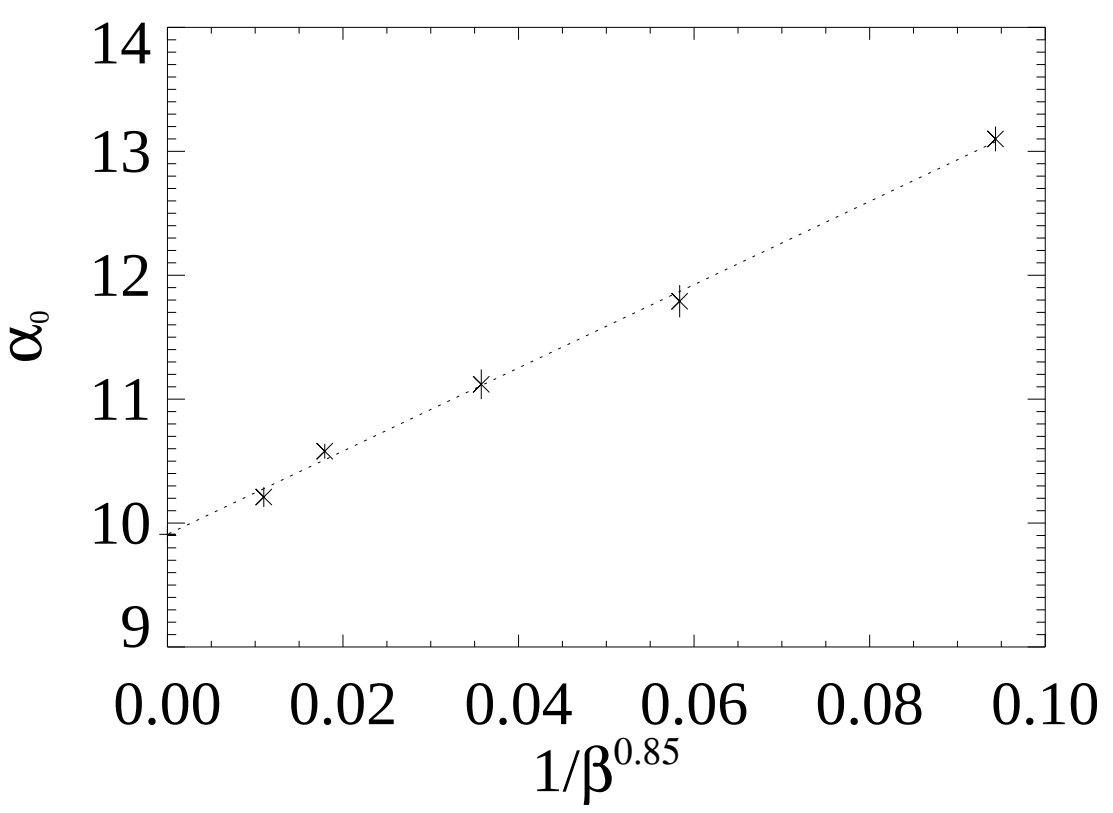

Figure 1. A linear extrapolation of $\alpha_0$ against $1/\beta^{0.85}$ to give the $\beta = \infty$ limit.

## 3. Qualitative Interface Behaviour Near $T_c$

At high $T \gg T_c$, the two $Z(3)$ phases introduced on the lattice by use of the twist are well defined and separated by a rigid interface; fluctuations in phase are very small. However, at and below $T_c$, these $Z(3)$ phases cease to exist, along with any $Z(3)$ interface, being replaced by the single disordered phase with symmetry restored. This raises the question: how does the interface behave between these two regimes?

As $T$ decreases towards $T_c$, a corresponding decrease is seen in the energy penalty suppressing phase fluctuations. One also expects the expectation values of Polyakov lines to converge on zero at the (second-order) phase transition, this being the value seen in the disordered phase. The combination of these factors means that the "height" of the $Z(3)$ interface (the difference in the real part of the Polyakov line as one moves across it) decreases just as more and larger bubbles of different phase appear within the two main $Z(3)$ domains on the lattice. The interface thus becomes increasingly difficult to discern as one approaches $T_c$. For our system, it is hard to see the interface on a plot of Polyakov line values below $\beta \approx 9.00$, against a critical value of $\beta_c \approx 8.175$[8]. This, of course, is precisely the temperature range of interest.

The first objective must be to uncover the qualitative behaviour of the interface as $T$ drops. Several possibilities present themselves, given its properties at high $T$ and at $T_c$: the interface could maintain its high-$T$ rigid structure with minimal "area", its "height" shrinking steadily until it vanishes at $T_c$; it could spread out longitudinally as correlation lengths on the lattice diverge with decreasing $T$, revealing more and more disordered phase and becoming less and less meaningful as a physical object; it could be washed out by the formation of many more interfaces along with the large bubbles of phase, colliding with each other and leaving disordered phase as the energy penalty drops; or it could maintain a relatively narrow width, but suffer increasingly violent transverse fluctuations which diverge towards $T_c$, where it would break apart to leave only disorder.

To follow the interface into the regime near $T_c$, one needs a way to identify it amidst the sea of phase fluctuations. One such way is to screen out the highest-frequency fluctuations, for instance by a straightforward box-car average over neighbouring Polyakov lines. This produces a spatial phase profile where the interface can be seen much more clearly, at the cost of some loss of information about its shape. The interface becomes visible as a localised, narrow object right down to $T_c$.

An improvement on this method is to contour the Polyakov profiles, picking various "heights" between the two extremes at opposite longitudinal ends of the lattice. Assuming no other interfaces are produced on the lattice, only one contour at each level will wrap across the lattice, having unit "winding number"; mere phase bubbles will be represented by "topologically trivial" contours. This produces a detailed picture of the phase structure of the interface at any stage of the simulation. Our results reveal that the interface does indeed remain thin and localised, but with increasingly large fluctuations in shape as $T$ drops, as illustrated in the snapshot of fig. 2. This matches the fourth of the possible behaviours listed previously.

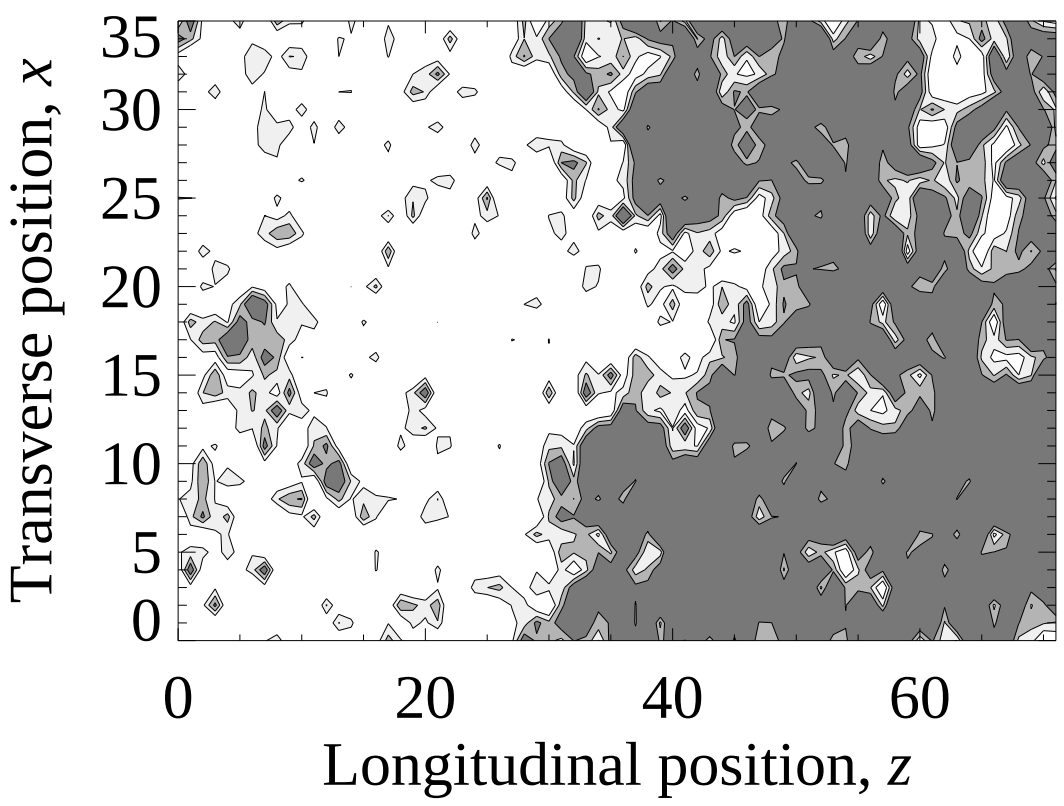

Figure 2. A contour snapshot of Polyakov lines (real parts) using three contour levels for $L_x = 36, L_z = 72$ and $\beta = 8.5$.

## 4. Quantitative Analysis

Having ascertained the qualitative behaviour of the interface to be that of a fluctuating string whose fluctuations increase as $T \to T_c$, the next task is to gain a better quantitative understanding. One might suspect that the fluctuations would diverge at $T_c$ itself, dissolving the interface and leaving only disorder, as desired. To test this, we follow various fluctuation moments: if $\phi(x)$ is the displacement from average of an interface contour at transverse location $x$, the $n$'th moment is defined to be $<\phi(x)^n>$. We track the first six moments, taking a separate average for each transverse position. Since the interface is translation invariant, with no preferred longitudinal position, one expects the moments to be roughly constant across the lattice after a large number of sweeps, and this is indeed what we see, with the even moments also found to be much larger than the odd. Thus, we can further average the moments across the lattice, and by producing such average moments for various temperatures near $T_c$, we see that the even moments do indeed diverge as $T \to T_c$.

The behaviour of the interface, rigid at high $T$ but with increasing fluctuations as $T$ drops, suggests a description in terms of a scalar field theory with a simple interaction to represent production and destruction of bubbles of phase at the interface. To maintain translation invariance, and

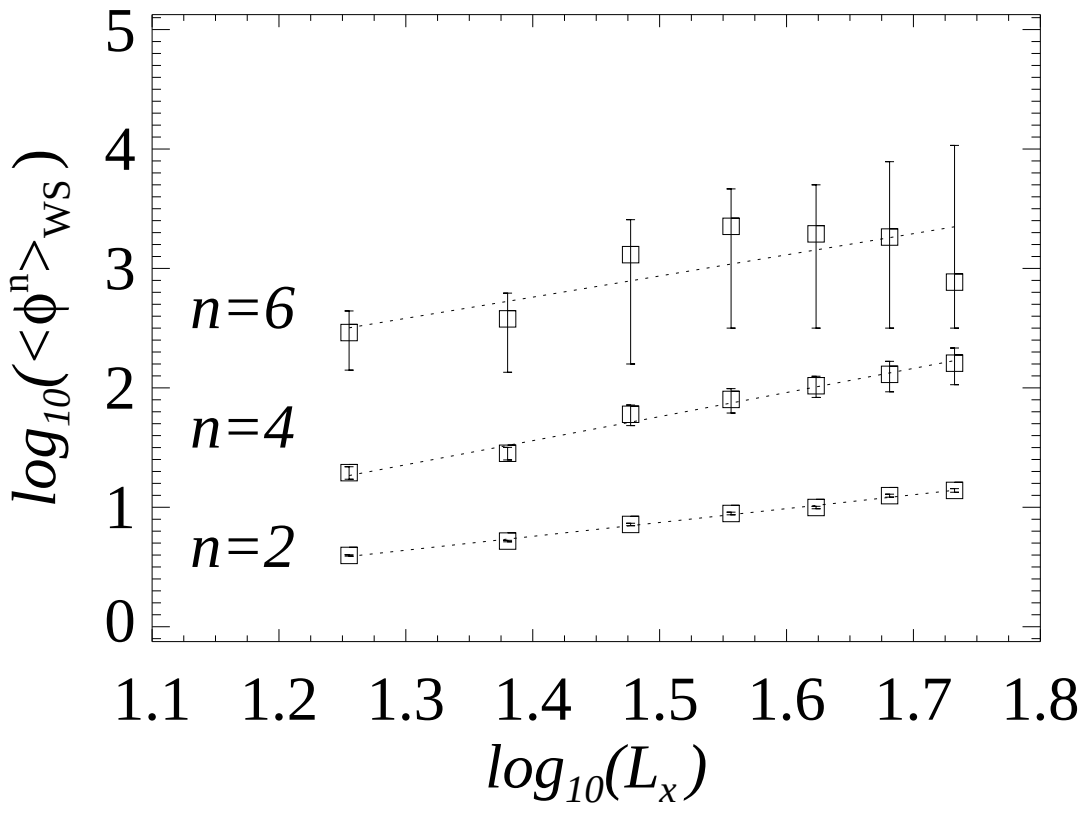

Figure 3. A log plot of the $x$-averaged, Wick-subtracted, $n$'th (even) moments as a function of $L_x$, for $\beta = 8.5$.

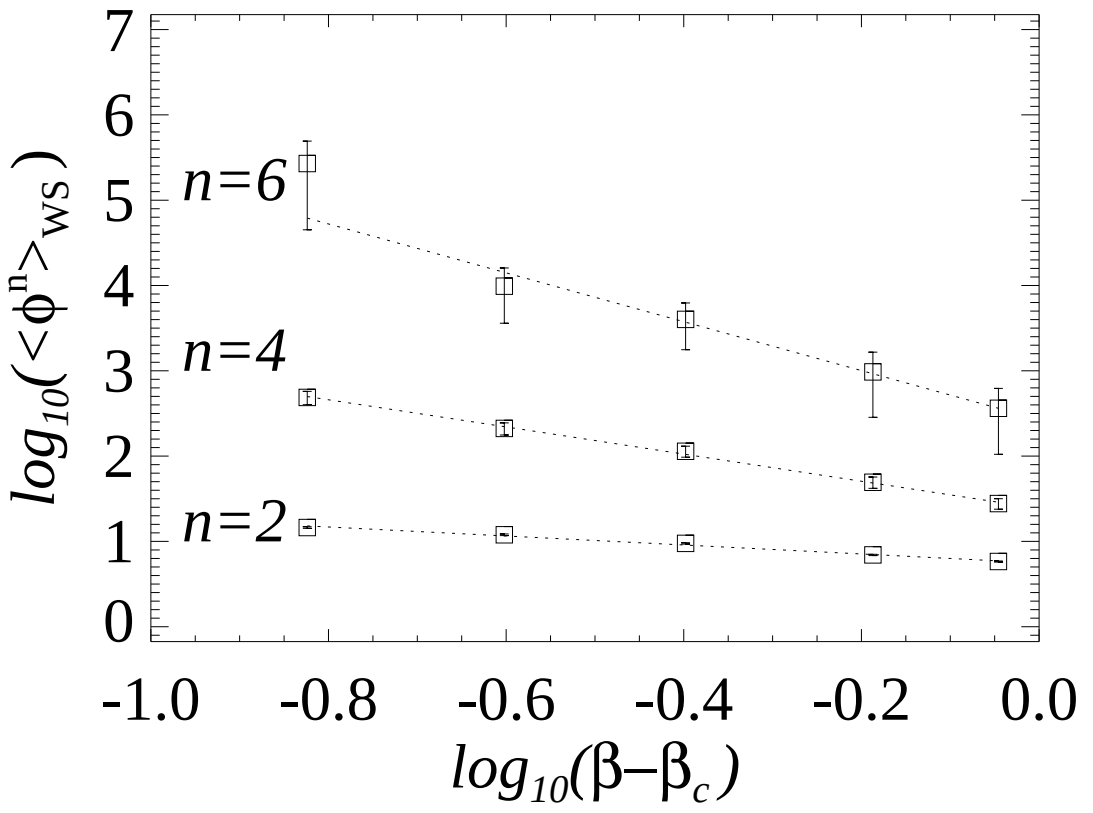

Figure 4. A log plot of the $x$-averaged, Wick-subtracted, $n$'th (even) moments as a function of $(\beta - \beta_c)$, for $L_x = 36$.

produce the domination of the even moments, we suggest a Lagrangian of the following form:

$$\mathcal{L} = \frac{1}{2}\gamma(\partial_x\phi)^2 - \frac{\lambda}{4!}(\partial_x\phi)^4. \quad (4)$$

Here, $\gamma$ and $\lambda$ are unknown functions of $(\beta - \beta_c)$. We then equate the $n$'th fluctuation moment, after Wick subtraction, to the $n$-point connected vacuum correlation function of this Lagrangian. As desired, the odd moments are predicted to be zero, since the interaction is of even order. A test of the applicability of the Lagrangian is provided by the dependence of the moments on the lattice width (the finite size of the string). Correlation functions predict dependences of $L_x$, $L_x^2$ and $L_x^3$ for the 2nd, 4th and 6th moments respectively, and this is roughly what we see in results such as fig. 3, with slopes of 1.16(1), 2.02(5) and 1.8(4).

Having satisfied ourselves that the scalar field theory model provides a reasonable description of the interface near its breakdown, we can measure the moments at different $T$ to find the functions $\gamma$ and $\lambda$. For $8.25 \leq \beta \leq 9.00$, we obtain results such as fig. 4. From the gradients, fitted to be -0.53(1), -1.59(3) and -2.9(4), we estimate that $\gamma \sim (\beta - \beta_c)^{0.53(1)}$ and $\lambda \sim (\beta - \beta_c)^{0.53(4)}$.

## 5. Conclusions

We have measured the tension of a $Z(3)$ interface in the $T = \infty$ limit, and found excellent agreement with theory, bolstering the argument for the existence of $Z(3)$-breaking phases in the Euclidean path integral. We have also studied the behaviour of this interface in a very different regime, near to its collapse. Measuring very close to the critical temperature, we have found its qualitative behaviour to be that of a string with divergent fluctuations at the critical point, these being well described by a simple interacting scalar field theory. Further analysis in continuing, and a similar study in 3+1 dimensions may well be rewarding.